\documentclass[letterpaper,11pt,leqno]{article}
\usepackage{paper,math}
\pdfoutput=1
\hypersetup{pdftitle={Has the Recession Started?}}
\available{https://pascalmichaillat.org/16/}
\newcommand{\bib}{paper.bib}
\newcommand{\pdf}{figures.pdf}
\begin{document}

\title{Has the Recession Started?}
\author{Pascal Michaillat, Emmanuel Saez
\thanks{Michaillat: University of California, Santa Cruz. Saez: University of California, Berkeley. We thank Megan Bailey, Austin Brown, Richard Crump, Brian Horrigan, Manfred Keil, Edward Nelson, Thomas Philips, Jack Schannep, Mike Shedlock, and Pawel Skrzypczynski for helpful comments.}}
\date{September 2024} 
\begin{titlepage}\maketitle

To answer this question, we develop a new Sahm-type recession indicator that combines vacancy and unemployment data. The indicator is the minimum of the Sahm indicator---the difference between the 3-month trailing average of the unemployment rate and its minimum over the past 12 months---and a similar indicator constructed with the vacancy rate---the difference between the 3-month trailing average of the vacancy rate and its maximum over the past 12 months. We then propose a two-sided recession rule: When our indicator reaches 0.3pp, a recession may have started; when the indicator reaches 0.8pp, a recession has started for sure. This new rule is triggered earlier than the Sahm rule: on average it detects recessions 0.8 month after they have started, while the Sahm rule detects them 2.1 months after their start. The new rule also has a better historical track record: it perfectly identifies all recessions since 1929, while the Sahm rule breaks down before 1960. With August 2024 data, our indicator is at 0.54pp, so the probability that the US economy is now in recession is 48\%. In fact, the recession may have started as early as March 2024.

\end{titlepage}\section{Introduction}

Has the US economy entered a recession? To answer the question, this note develops a new \citet{S19}-type recession indicator that combines data on job vacancies and unemployment.

The indicator is the minimum of the Sahm indicator---the difference between the 3-month trailing average of the unemployment rate and its minimum over the past 12 months---and a similar indicator constructed with the vacancy rate---the difference between the 3-month trailing average of the vacancy rate and its maximum over the past 12 months. It then proposes a new recession rule: a recession may have started when the minimum indicator reaches 0.3pp.

This new rule is triggered earlier than the \citet{S19} rule---which only uses unemployment data. It detects recessions with a lag of 0.8 month on average, while the Sahm rule detects them with a lag of 2.1 months. The new rule also has a better historical track record. It detects all recessions since 1929, without any false positives, while the Sahm indicator breaks down before 1960.

A one-sided recession rule such as the Sahm rule tells us whether a recession might have started or not. To know what is the likelihood that a recession has started, we propose a two-sided recession rule. The bottom threshold is the lowest value that generates no false positives over 1960--2022 (this is how the threshold of 0.5pp is determined in the Sahm rule). The top threshold is the highest value that generates no false negatives over 1960--2022. The two-sided rule is as follows. When the minimum indicator is below 0.3pp, we cannot say that the recession has started. When the minimum indicator is between 0.3pp and 0.8pp, the recession may have started. And when the minimum indicator is above 0.8pp, the recession has started for sure.

Our recession indicator crossed 0.3pp between March 2024 and April 2024, so the US economy might have entered a recession then. With August 2024 data, the minimum indicator is at 0.54pp, so the probability that the US economy is now in recession is $(0.54-0.3)/(0.8-0.3) = 48\%$. 

\section{Construction of the real-time recession indicator}

In this section we construct our real-time recession indicator by combining unemployment and vacancy data for the United States, January 1960--August 2024.

\subsection{Data on unemployment, job vacancies, and recessions}

\begin{figure}[t]
\includegraphics[scale=\wscale,page=1]{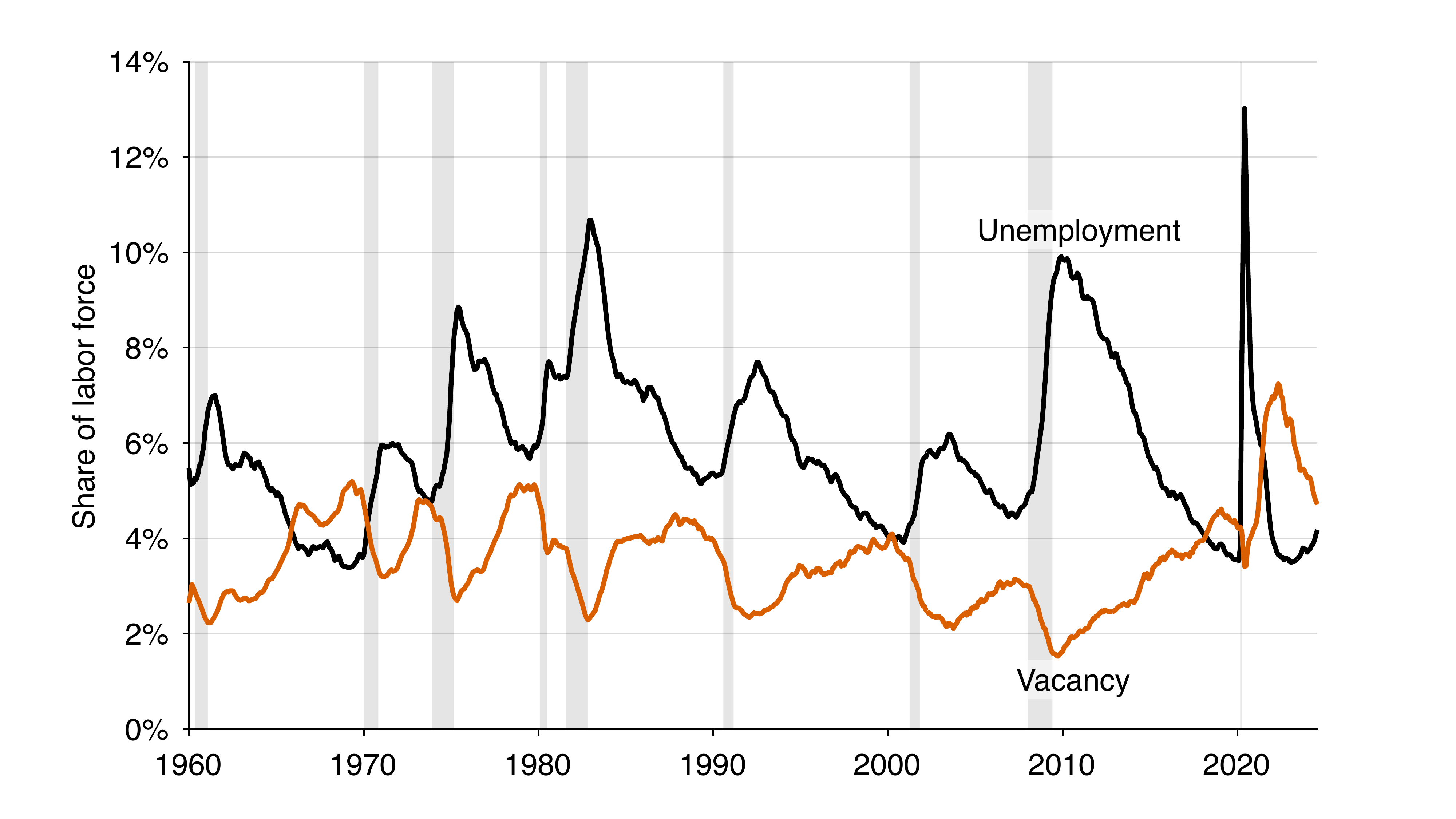}
\caption{Unemployment and vacancy rates in the United States, January 1960--August 2024}
\note{The unemployment rate is computed from data produced by the \citet{UNEMPLOY,CLF16OV}. The vacancy rate is computed from data produced by \citet{B10} and the \citet{CLF16OV,JTSJOL}. The unemployment and vacancy rates are 3-month trailing averages of monthly series. The gray areas are recessions dated by the \citet{NBER23}. }
\label{f:data}\end{figure}

The unemployment rate $u$, vacancy rate $v$, and recession dates that we use in the analysis are plotted in figure~\ref{f:data}. These unemployment and vacancy data are widely used in macroeconomics \citep{DHS12,DS15,EMR15,BF15,BFH24,MS21b,MS24}.

\paragraph{Unemployment rate}  The unemployment rate is the number of jobseekers measured by the \citet{UNEMPLOY} from the Current Population Survey (CPS), divided by the civilian labor force constructed by the \citet{CLF16OV} from the CPS. This is the standard, official measure of unemployment, labelled U3 by the \citet{BLS23}.\footnote{The Sahm indicator is constructed with the unemployment rate produced by the \citet{UNRATE}, which takes the same values as our unemployment rate but is rounded to the first digit \citep{SAHMCURRENT}. The rounding unnecessarily adds volatility to the indicator, which is especially problematic in the vicinity of the recession threshold. For instance, in July 2024 the Sahm indicator crossed its 0.5pp threshold with the rounded unemployment rate (0.53pp), but not with the unrounded unemployment rate (0.49pp). So here, we simply use the exact, unrounded unemployment rate.} 

\paragraph{Vacancy rate} The vacancy rate is derived from two different sources because there is no continuous vacancy series over the period. For 1960--2000, we use the vacancy rate constructed by \citet{B10}. This series is based on the Conference Board's help-wanted advertising index, adjusted to account for the shift from print advertising to online advertising in the 1990s. The Conference Board index aggregates help-wanted advertising in major metropolitan newspapers in the United States. It serves as a reliable proxy for job vacancies \citep{A87,S05}. For 2001--2024, we use the number of job openings measured by the \citet{JTSJOL} from the Job Openings and Labor Turnover Survey (JOLTS), divided by the civilian labor force constructed by the \citet{CLF16OV} from the CPS. To best align labor force and vacancy data, we follow \citet{MS24} and shift forward by one month the number of job openings from JOLTS.\footnote{For instance, we assign to December 2023 the number of job openings that the BLS assigns to November 2023. The motivation for this shift is that the number of job openings from the JOLTS refers to the last business day of the month (Thursday 30 November, 2023), while the labor force from the CPS refers to the Sunday--Saturday week including the 12th of the month (Sunday 10 December 2023 to Saturday 16 December 2023) \citep{BLS20,BLS24}. So the number of job openings refers to a day that is closer to next month's CPS reference week than to this month's CPS reference week.} We then splice the two series to create a continuous vacancy rate for January 1960--August 2024. The two series are perfectly aligned because \citet{B10} used the JOLTS data to scale the Conference Board index and translate it into a vacancy rate (which was possible because the Conference Board and JOLTS series overlap in the early 2000s).

\paragraph{Recession dates} The \citet{NBER23} identifies the peaks and troughs of US business cycles. Following their convention, we set the first month of the recession as the month following the peak and the last month of the recession as the month of the trough \citep{NBER22}. 

\paragraph{Data availability} The data required to construct our indicator for any given month are released on the first week of the following month, usually on a Tuesday for the JOLTS data and a Friday for the CPS data \citep{BLS24b}.\footnote{This is another advantage of shifting the number of job openings from the JOLTS forward by one month. We have access to the vacancy and unemployment rates required to compute our real-time indicator on the same week, as soon as the month is over. The CPS data are communicated as part of the "Employment Situation" release.} So the indicator can be constructed in real time. The number of job openings released by the BLS is preliminary and is updated one month after its first release. Therefore, the real-time value of indicator is preliminary, before it receives its final update on month later. By contrast, the official dates of recessions are published after a long and variable delay by the Business Cycle Dating Committee of the \citet{NBER21}.

\subsection{Construction of the recession indicator}

We start from the \citet{S19} recession indicator. That indicator is computed in two steps. The first step is taking the 3-month trailing average of the unemployment rate. The second step is taking the difference between the average unemployment rate and its minimum over the past 12 months. 

Formally, we denote the monthly unemployment rate by $u(t)$. The first step produces the 3-month trailing average:
\begin{equation*}
\bar{u}(t) = \frac{u(t)+u(t-1)+u(t-2)}{3}.
\end{equation*}
The second step takes the difference between the trailing average and its 12-month trailing minimum:
\begin{equation}
\hat{u}(t) = \bar{u}(t) - \min[0\leq s \leq 12]{\bar{u}(t-s)}.
\label{e:uindicator}\end{equation}
The variable $\hat{u}(t)$ is the unemployment indicator. The unemployment indicator is always positive: it is zero when unemployment is decreasing but it turns strictly positive once unemployment starts rising.\footnote{The standard Sahm indicator is sometimes negative \citep{SAHMCURRENT}. This is because the trailing minimum is taken over the previous 12 months without including the current month: $\hat{u}(t) = \bar{u}(t) - \min[1\leq s \leq 12]{\bar{u}(t-s)}$ \citep{S23}. By contrast, our indicators are always nonnegative, which is neater without affecting the results.}

Our indicator is based on the same idea, but it combines vacancy and unemployment data to be able to detect recessions more quickly and more robustly. Indeed, job vacancies start falling quickly at the onset of recessions, when unemployment starts rising (figure~\ref{f:data}). Requiring that both rise gives a more accurate and---maybe counterintuitively---more rapid recession signal. 

We therefore construct a vacancy indicator by taking the 3-month trailing average of the vacancy rate, and then by taking the difference between the average vacancy rate and its maximum over the past 12 months. Formally, we denote the monthly vacancy rate by $v(t)$. The first step produces the 3-month trailing average:
\begin{equation*}
\bar{v}(t) = \frac{v(t)+v(t-1)+v(t-2)}{3}.
\end{equation*}
The second step takes the difference between the trailing average and its 12-month maximum:
\begin{equation}
\hat{v}(t) = \max[0\leq s \leq 12]{\bar{v}(t-s)} - \bar{v}(t).
\label{e:vindicator}\end{equation}
The variable $\hat{v}(t)$ is the vacancy indicator. It is always positive: zero when vacancies are increasing but strictly positive once vacancies start falling.

Finally, the minimum indicator that we will use to identify recession starts is the minimum of the two previous indicators:
\begin{equation}
x(t) = \min{\hat{u}(t), \hat{v}(t)}.
\label{e:minindicator}\end{equation}

\subsection{Construction of the recession rule}

Using the unemployment and vacancy data from figure~\ref{f:data}, we compute an unemployment indicator using formula \eqref{e:uindicator} (black line in figure~\ref{f:uv}). Then we construct the vacancy indicator using \eqref{e:vindicator} (orange line in figure~\ref{f:uv}). Figure~\ref{f:uv} also plots the threshold that \citet{S19} proposes to identify recession starts: 0.5pp. The Sahm rule, which combines the unemployment indicator with a threshold of 0.5pp, works perfectly between January 1960 and December 2022. It has no false positives (detected recessions that are not actual recessions) and no false negatives (actual recessions that are not detected as recessions).\footnote{We stop the evaluation of the recession rule at the end of 2022 because it is too early to say if a recession officially occurred in 2023--2024. Indeed, the \citet{NBER21} may take as long as one year to date the official start of a recession.}

In July 2024, the unemployment indicator was just below the 0.5pp threshold, at 0.49pp. But the unemployment indicator then crossed the 0.5pp threshold, reaching 0.54pp in August 2024. The Sahm rule therefore says that as of August 2024, the United States has just entered a recession.

\begin{figure}[p]
\subcaptionbox{Unemployment and vacancy indicators\label{f:uv}}{\includegraphics[scale=\wscale,page=2]{\pdf}}\\
\subcaptionbox{Minimum indicator\label{f:minimum}}{\includegraphics[scale=\wscale,page=3]{\pdf}}
\caption{Recession indicators in the United States, January 1960--August 2024}
\note{The unemployment indicator is computed with \eqref{e:uindicator}. The vacancy indicator is computed with \eqref{e:vindicator}. The minimum indicator is computed with \eqref{e:minindicator}. The unemployment and vacancy rates used to compute the indicators come from figure~\ref{f:data}. The gray areas are NBER-dated recessions. The unemployment indicator was proposed by \citet{S19}.}
\label{f:indicators}\end{figure}

To get a better sense of whether a recession has already started, we need a faster recession rule, that predicts recessions earlier. To have a faster recession rule, we need to be able to lower the 0.5pp threshold without generating false positives. But this is not possible with the unemployment indicator because in June 2003 the unemployment indicator reached 0.5pp but there was no recession (figure~\ref{f:uv}).

Looking at figure~\ref{f:uv}, the vacancy indicator appears broadly as fast as the unemployment indicator. It would call some recessions slightly earlier (in 1990 and 2001) and some recessions slightly later (in 2008). In the aftermath of the pandemic, the vacancy indicator started rising in 2022 and peaked in 2023, so it would have delivered a prediction that was so early as to be misleading. This might be due to the extreme outward shift of the Beveridge curve during the pandemic \citep{MS24}. This shift leads to elevated values of the vacancy rate during that period, and therefore elevated values of the vacancy indicator.

But the main advantage of the vacancy indicator is that it does not present the same uninformative blips as the unemployment indicator. For instance, there is no problematic blip in June 2003: the vacancy indicator is not 0 but it is much lower than the unemployment indicator. Of course, it presents other uninformative blips. For instance it has a peak in July 1967 while there was no recession then.\footnote{Friedman and Schwartz did argue that a minirecession occurred in 1966--1967 \citep[pp. 102--110]{N20b}.}

To have a more accurate recession rule, we therefore use an indicator that is the minimum of the unemployment and vacancy indicators. Given that the blips of the unemployment and vacancy indicators do not occur at the same time, taking the minimum of the two indicators will eliminate these blips, and given us a less noisy, more accurate recession indicator. The reduced noise will then allow us to lower the detection threshold below 0.5pp. Of course the minimum of the two indicators will be slower to increase---since it can only increase when both indicators rise. But the reduction in threshold afforded by the reduced noise will be so large that the minimum indicator will detect recession faster on average.

The minimum indicator, constructed from formula \eqref{e:minindicator}, is plotted on figure~\ref{f:minimum}. Because the blips from the unemployment and vacancy indicators are eliminated, we can lower the threshold to call recessions to 0.3pp. We could not lower the threshold below 0.3pp because in 2003 the minimum indicator reached 0.3pp but there was no recession.

\begin{figure}[t]
\includegraphics[scale=\wscale,page=8]{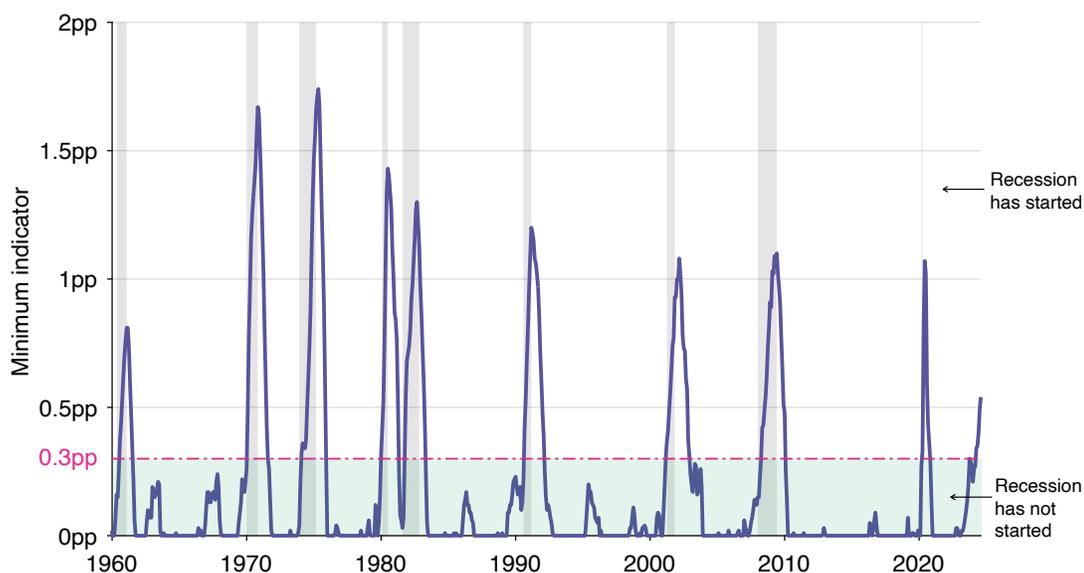}
\caption{Minimum indicator and one-sided recession rule in the United States, January 1960--August 2024}
\note{The minimum indicator is computed with \eqref{e:minindicator}. The unemployment and vacancy rates used to compute the indicator come from figure~\ref{f:data}. The gray areas are NBER-dated recessions. When the indicator is below 0.3pp, we cannot say that a recession has started. When the indicator is above 0.3pp, a recession has started.}
\label{f:onesided}\end{figure}

The recession rule obtained by combining our minimum indicator with a recession threshold of 0.3pp works perfectly between January 1960 and December 2022 (figure~\ref{f:onesided}). First, the recession rule has no false positives---detected recessions that are not actual recessions. Second, the recession rule has no false negatives---actual recessions that are not detected as recessions. 

\section{Detected recessions and detection delay}\label{s:delay}

Our recession rule---based on the minimum indicator and a threshold of 0.3pp---has a prefect track record over January 1960--December 2022. This is just like the \citet{S19} rule---which is based on the unemployment indicator and a threshold of 0.5pp. The two rules identify the 9 recessions that occurred during the period, without any false positives (figure~\ref{f:indicators}). The stories behind these recessions are well known:
\begin{itemize}
\item The recession that started in March 2020 was caused by the coronavirus pandemic.
\item The recession that started in January 2008 coincided with the global financial crisis.
\item The recession that started in April 2001 followed the burst of the dot-com bubble.
\item The recession that started in August 1990 followed the Iraqi invasion of Kuwait and associated oil price shock.
\item The recessions that started in February 1980 and August 1982 are associated with the Volcker disinflation's tight monetary policy.
\item The recession that started in December 1973 followed the (first) oil crisis.
\item The recession that started in January 1970 coincided with fiscal and monetary tightening toward the end of the Vietnam War.
\item The recession that started in May 1960 followed tighter monetary policy in 1958--1960.
\end{itemize}

\begin{table}[t]
\caption{Recession start dates in the United States, January 1960--December 2022}
\begin{tabular*}{\textwidth}{@{\extracolsep{\fill}}c*{5}{c}}
\toprule
\multicolumn{2}{c}{Official start dates} & \multicolumn{2}{c}{Unemployment indicator $>$ 0.5pp}  & \multicolumn{2}{c}{Minimum indicator $>$ 0.3pp} \\
\cmidrule{1-2}\cmidrule{3-4}\cmidrule{5-6}
Year   &  Month   & Year   &  Month & Year   &  Month  \\
\midrule
1960   &  5   & 1960   &    9 & 1960   &    7  \\
1970   &  1   & 1970   &    2 & 1970   &     1  \\
1973   & 12   & 1974   &    6 & 1974   &     2  \\
1980   &  2   & 1980   &    2 & 1980   &     1  \\
1981   &  8   & 1981   &    11 & 1981   &    10  \\
1990   &  8   & 1990   &    10 & 1990   &    8  \\
2001   &  4   & 2001   &    6 & 2001   &     3  \\
2008   &  1   & 2008   &    2 & 2008   &     4  \\
2020   &  3   & 2020   &    3 & 2020   &     3  \\
\midrule
\multicolumn{2}{c}{Average detection delay:} & \multicolumn{2}{c}{2.1 months}  & \multicolumn{2}{c}{0.8 month} \\
\bottomrule\end{tabular*}
\note{The unemployment and minimum indicators and the thresholds are displayed in figure~\ref{f:indicators}. The official start dates are provided by the \citet{NBER23}. The start dates detected by the Sahm rule (unemployment indicator $>$ 0.5pp) are computed with \eqref{e:datingSahm}. The start dates detected by our recession rule (minimum indicator $>$ 0.3pp) are computed with \eqref{e:dating}.}
\label{t:delay}\end{table}

Whenever the minimum indicator crosses its threshold, $x(t)>0.3$pp, our recession rule says that the US economy may have entered a recession (figure~\ref{f:onesided}). To date the start of the recession, we assess whether the crossing occurred in the current month or the previous month. If the current value of the indicator is closer to the threshold than the previous value, we set the recession start to the current month. But if the value of the indicator in the previous month is closer to the threshold than the current value,  we set the recession start to the previous month. Formally, if the indicator crossed the threshold between months $t-1$ and $t$, we set the recession start month to
\begin{equation}
t - \ind{0.003 - x(t-1) < x(t) - 0.003}.
\label{e:dating}\end{equation}

The logic behind the dating rule \eqref{e:dating} is that the indicator is evolving continuously over time. The unemployment and labor force numbers describe labor market conditions roughly in the middle of each month.\footnote{The CPS reference week, during which the CPS survey is conducted, is the Sunday--Saturday week including the 12th of the month \citep{BLS20}.} The number of job openings describe conditions at the very beginning of the month.\footnote{Technically, the number of job openings refer to the last day of the previous month \citep{BLS24}.} To be conservative, we assign the conditions described by the indicator to the middle of each month. Then, the threshold was crossed in the current month only if the threshold is closer to the current month's value of the indicator than to the previous month's value of the indicator. 

Before comparing the recession start dates generated by our recession rule to those generated by the Sahm rule, we compute the recession start month from the Sahm rule using a similar dating rule:
\begin{equation}
t - \ind{0.005 - \hat{u}(t-1) < \hat{u}(t) - 0.005}.
\label{e:datingSahm}\end{equation}

An added advantage of the dating rules \eqref{e:dating} and \eqref{e:datingSahm} is that they increase the informativeness of the start dates produced by the recession rules. This is easy to see by considering the discussion around the Sahm rule in July--August 2024. Based on unrounded unemployment rates, the unemployment indicator reached 0.49pp in July and then 0.54pp in August. So the Sahm rule was only triggered in August, but the dating rule \eqref{e:datingSahm} then ascribes the recession start to July, since the July value of the indicator was so close to the threshold. In that way, the start date would remain the same whether the July value was 0.47pp or 0.48pp or 0.49pp or 0.50pp or 0.51pp. What matters for the start date is the closeness to the threshold, not whether the indicator is just above or just below it.

We find that our recession rule is able to identify recessions faster than the Sahm rule (table~\ref{t:delay}). Our recession rule detects recession starts with a delay of 0.8 month compared to the official recession starts determined by the \citet{NBER23}. This is more than a month faster than the Sahm rule, which detects recession starts with a delay of 2.1 months compared to the official recession starts. Our recession rule is always faster than the Sahm rule, except in 2008 when it called the Great Recession 2 months later than the Sahm rule (in April 2008 instead of February 2008). The slight delay is because job vacancies took some time to drop at the onset of the Great Recession (the delay is visible in figures~\ref{f:data} and \ref{f:uv}).

Of course, it is not surprising that the recession starts obtained with these indicators often lag the NBER recession starts, because the NBER recession starts are backdated \citep{NBER21}. The NBER dates are identified with hindsight, not in real time, which is what our indicators try to do.

\section{Current recession probability}\label{s:recession_probability}

As we know, and as we can see on figure~\ref{f:uv}, the unemployment indicator crossed 0.5pp between July and August 2024, implying through the Sahm rule that a recession might have started then.

\begin{figure}[p]
\includegraphics[scale=\wscale,page=4]{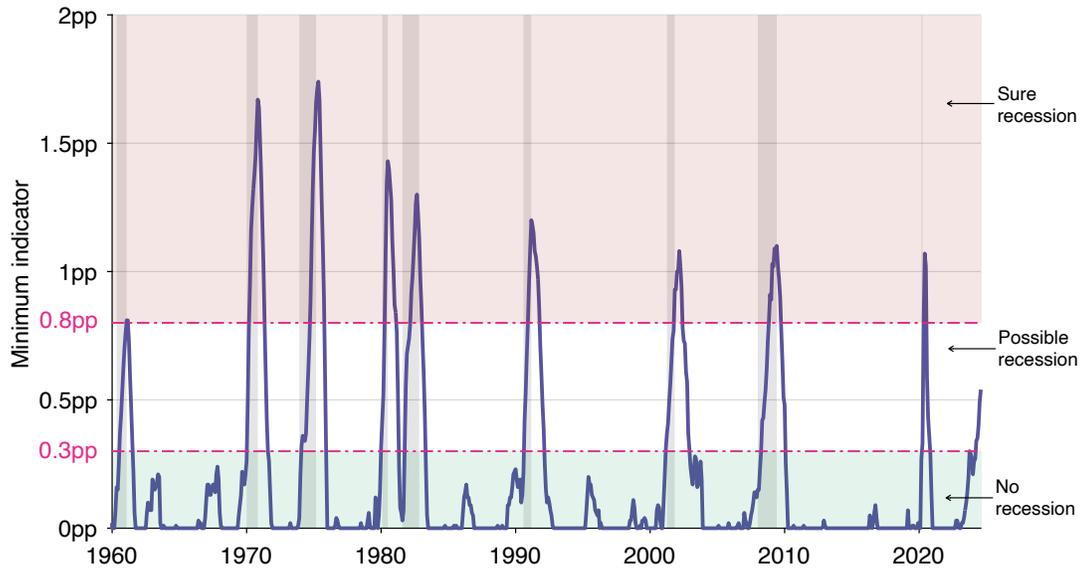}
\caption{Minimum indicator and two-sided recession rule in the United States, January 1960--August 2024}
\note{The minimum indicator is computed with \eqref{e:minindicator}. The unemployment and vacancy rates used to compute the indicator come from figure~\ref{f:data}. The gray areas are NBER-dated recessions. When the indicator is below 0.3pp, we cannot say that a recession has started. When the indicator is above 0.8pp, a recession has started for sure. When the indicator is in the 0.3pp--0.8pp band, a recession is likely to have started.}
\label{f:twosided}\end{figure}

What does our recession rule say? The minimum indicator actually crossed the 0.3pp threshold between March and April 2024, indicating that a recession might have started a few months ago (figure~\ref{f:twosided}). It is not surprising that our recession rule is able to call the current recession earlier than the Sahm rule, since our recession rule is systematically faster than the Sahm rule (table~\ref{t:delay}). In August 2024, the minimum indicator reached 0.54pp, so it is well above the recession threshold of 0.3pp.

\begin{figure}[p]
\includegraphics[scale=\wscale,page=5]{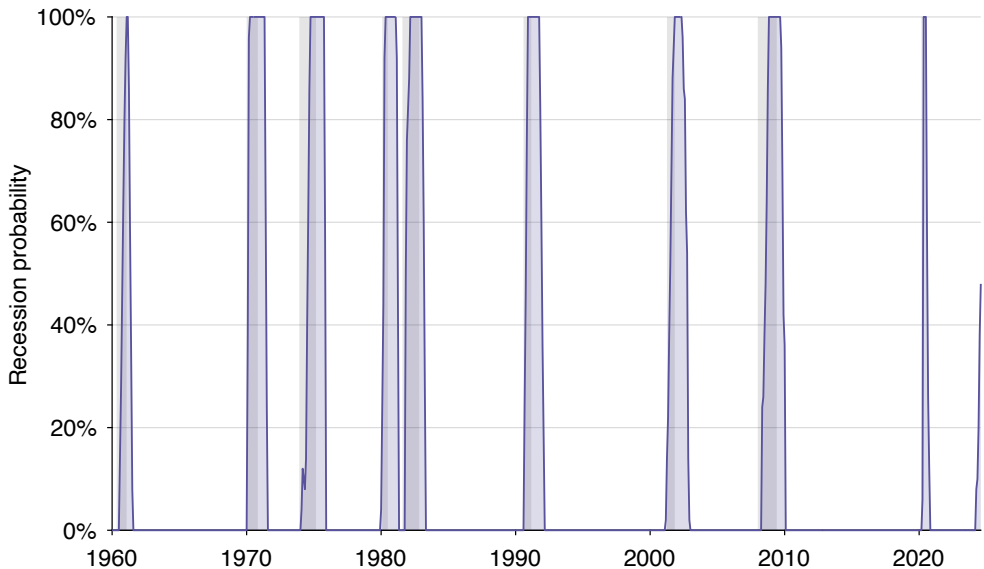}
\caption{Recession probability in the United States, January 1960--August 2024}
\note{The recession probability is given by formula \eqref{e:probability}. The formula uses the minimum indicator and thresholds displayed in figure~\ref{f:twosided}. The gray areas are NBER-dated recessions.}
\label{f:probability}\end{figure}

In fact, we can add a second threshold to our rule so as to compute the probability that the US economy has entered a recession. The current debate about whether the Sahm rule has been triggered or not, and whether the US economy has entered a recession or not, comes from the fact that the current analysis only uses one threshold. This threshold is the lowest threshold such that the rule does not produce false positives between January 1960 and December 2022. This threshold is 0.5pp in the case of the unemployment indicator (Sahm rule), and 0.3pp in the case of the minimum indicator. The threshold cannot be lowered below 0.3pp because the minimum indicator reached 0.28pp in 2003 while there was no recession then.

But we can also produce another threshold, which is the highest threshold such that the indicator does not produce false negatives between January 1960 and December 2022. In the case of the minimum indicator, this threshold is 0.8pp (figure~\ref{f:twosided}). This threshold, which is the highest value that the indicator can reach before we are bound to acknowledge that the economy is in a recession. This conservative threshold cannot be raised above 0.8pp because the minimum indicator reached 0.81pp in 1960, which was a recession year.

With the two thresholds, 0.3pp and 0.8pp, we have a two-sided rule. When the minimum indicator is below 0.3pp, we cannot say that the US economy is in a recession. When the indicator is above 0.8pp, we must acknowledge that the US economy is in a recession. When the indicator is between 0.3pp and 0.8pp, the two-sided rule says that a recession may have started.

Furthermore, we can compute the probability that the economy is in recession at any point when the indicator is between 0.3pp and 0.8pp. The probability simply reflects the share of the 0.3pp--0.8pp band that has been covered by the indicator. For instance, in August 2024, the minimum indicator is 0.54pp, so the probability that the US economy is in a recession is $(0.54-0.3)/(0.8-0.3) = 48\%$ (figure~\ref{f:probability}). In general, when the minimum indicator has a value of $x(t)\in [0.3,0.8]$, the probability that the recession has started is
\begin{equation}
p(t) = \frac{x(t) - 0.3}{0.8-0.3}.
\label{e:probability}\end{equation}

The recession probability is a byproduct of our ignorance, itself caused by the dearth of macroeconomic data. We start from the presumption that there is a unique threshold separating recessions from non-recessions. Any time our indicator crosses the true threshold from below, the economy enters a recession. The challenge is that there is not enough data to identify this threshold with exactitude. We know that the threshold is above 0.3pp because the indicator has crossed all the values below 0.3pp without triggering a recession. We also know that the indicator is below 0.8pp because there are recessions in the January 1960--December 2022 period that have not strictly crossed 0.8pp. So the latent threshold must be between 0.3pp and 0.8pp. We cannot narrow the range further because we have not observed more recessions. Assuming that this unobservable recession threshold is uniformly distributed over 0.3pp--0.8pp, we compute the probability to be in a recession as the probability that the indicator has crossed the latent threshold. This probability is given by formula \eqref{e:probability}.

\section{Historical track record}\label{s:record}

Next, we examine the historical performance of our recession rule. We extend the analysis to April 1929--December 1959, so that we cover almost one hundred years of US business cycles. Our recession rule continues to perform well in the past: it identifies all the recessions between April 1929 and December 1959 without any false positives.

\subsection{Historical data}

\paragraph{Unemployment rate} For April 1929--December 1947, the monthly unemployment rate is constructed by \citet{PZ21}. They extrapolate \citet{W92a}'s annual unemployment series to a monthly series using monthly unemployment rates compiled by the National Bureau of Economic Research (NBER). For January 1948--December 1959, the unemployment rate is computed just as in the modern period: it is the number of jobseekers divided by the civilian labor force, both measured by the \citet{UNEMPLOY,CLF16OV} from the CPS.

\paragraph{Vacancy rate} For April 1929--December 1950, the vacancy rate is based on help-wanted index created by the Metropolitan Life Insurance Company. This index aggregates help-wanted advertisements from newspapers across major US cities. It is considered a reliable proxy for job vacancies \citep{Z98a}. The MetLife index is scaled to align with \citet{B10}'s vacancy rate at the end of 1950, effectively translating the index into a vacancy rate.\footnote{\citet{PZ21} produce a vacancy series that starts in 1919 and an unemployment series that starts in 1890. We begin our analysis in April 1929, however, because there are some limitations with the prior data \citep[section 3B]{MS24}.} For January 1951--December 1959, we use again the vacancy rate produced by \citet{B10}.

\subsection{Historical detection of recessions}

Our recession rule continues to perform well in the past. Using a threshold of 0.3pp to detect the start of a recession, the minimum indicator identifies the 6 recessions of the April 1929--December 1959 period without producing any false positive (figure~\ref{f:record1}). That historical period featured extreme macroeconomic volatility, with vast fluctuations in unemployment and job vacancies, driven by seismic events such as the Great Depression and World War 2 \citep[section 3B]{MS24}. Nevertheless, our recession rule continues to work.

One difference in the historical period is that our recession rule is somewhat slower at detecting recessions starts. Over the April 1929--December 1959 period, on average, the  rule detects recession starts with a delay of 2.5 months compared to the official recession starts determined by the \citet{NBER23} (table~\ref{t:delay1929}). This is slower than over the modern period, when our rule's detection delay is only 0.8 month (table~\ref{t:delay}). A possible explanation for the delay is that the unemployment and vacancy data for that period might be noisier than the modern data. This noisiness might itself be explained by the fact that the unemployment data come from a patchwork of sources, and that the vacancy data were collected by private entities and not by the BLS \citep{PZ21}.

\begin{table}[t]
\caption{Recession start dates in the United States, April 1929--December 1959}
\begin{tabular*}{\textwidth}{@{\extracolsep{\fill}}c*{3}{c}}
\toprule
\multicolumn{2}{c}{Official start dates} & \multicolumn{2}{c}{Minimum indicator $>$ 0.3pp} \\
\cmidrule{1-2}\cmidrule{3-4}
Year   &  Month   & Year   &  Month  \\
\midrule
1929     &     9    &    1930   &    2\\
1937     &     6    &    1937   &   11\\
1945     &     3    &    1945   &    8\\
1948     &    12    &    1948   &   12\\
1953     &     8    &    1953   &    10\\
1957     &     9    &    1957   &    7\\
\midrule
\multicolumn{2}{c}{Average detection delay:} & \multicolumn{2}{c}{2.5 months} \\
\bottomrule\end{tabular*}
\note{The minimum indicator and threshold are displayed in figure~\ref{f:record1}.  The official start dates are provided by the \citet{NBER23}. The start dates detected by our recession rule (minimum indicator $>$ 0.3pp) are computed with \eqref{e:dating}.}
\label{t:delay1929}\end{table}

Over the historical period, it is necessary to adjust the two-sided recession rule, however. Because of the recession starting in 1937, it is necessary to lower the top threshold to avoid producing a false negative. The highest possible threshold that does not produce any false negatives over April 1929--December 2022 is 0.6pp (figure~\ref{f:record2}). With this lower top threshold, the probability that the US economy has entered a recession in August 2024 would climb to $(0.54-0.3)/(0.6-0.3) = 80\%$.

\begin{figure}[p]
\subcaptionbox{Minimum indicator and one-sided recession rule\label{f:record1}}{\includegraphics[scale=\wscale,page=9]{\pdf}}\\
\subcaptionbox{Minimum indicator and two-sided recession rule \label{f:record2}}{\includegraphics[scale=\wscale,page=6]{\pdf}}
\caption{Historical record of the minimum indicator in the United States, April 1929--August 2024}
\note{The minimum indicator is computed with \eqref{e:minindicator}. The unemployment rate used to compute the indicator come from \citet{PZ21} and \citet{UNEMPLOY,CLF16OV}. The vacancy rate used to compute the indicator come from \citet{PZ21}, \citet{B10}, and \citet{CLF16OV,JTSJOL}. The gray areas are NBER-dated recessions. When the minimum indicator is below 0.3pp, we cannot say that a recession has started. When the minimum indicator is above 0.6pp, a recession has started for sure. When the minimum indicator is in the 0.3pp--0.6pp band, a recession is likely to have started.}
\label{f:record}\end{figure}

\begin{figure}[t]
\includegraphics[scale=\wscale,page=7]{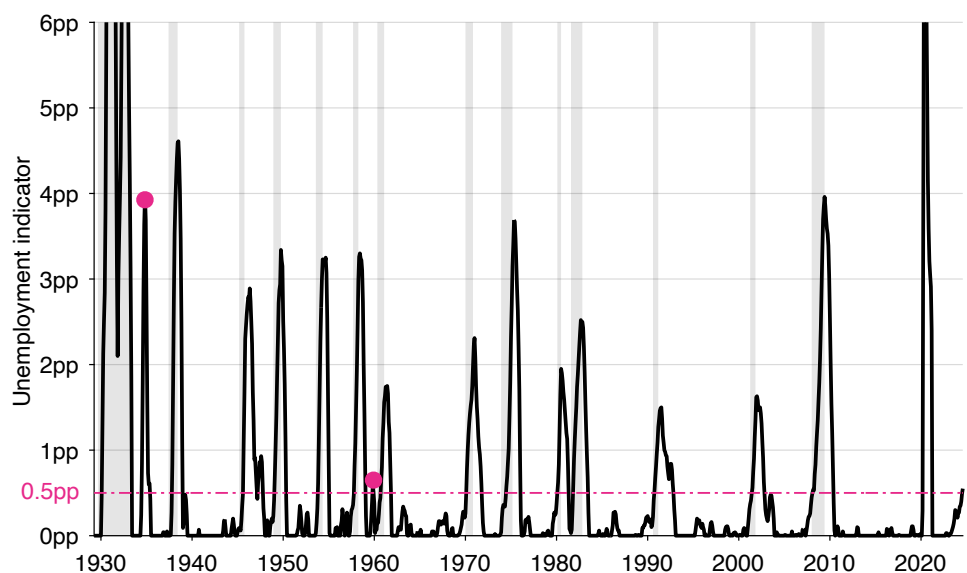}
\caption{Historical record of the Sahm rule in the United States, April 1929--August 2024}
\note{The unemployment indicator is computed with \eqref{e:uindicator}. The unemployment rate used to compute the unemployment indicator come from \citet{PZ21} and \citet{UNEMPLOY,CLF16OV}. The gray areas are NBER-dated recessions. The Sahm rule is that a recession starts when the indicator reaches 0.5pp.}
\label{f:recordSahm}\end{figure}

The Sahm rule works perfectly for January 1960--December 2022. But it breaks down just before 1960 because in 1959 the unemployment indicator reached 0.6pp although there was no recession (figure~\ref{f:recordSahm}). This issue could easily be fixed by raising the threshold used in the Sahm rule to 0.6pp. This would make the rule a little slower at detecting recession starts, but it would allow the rule to continue working until World War 2.

Before World War 2, the Sahm rule faces a much bigger problem and breaks down (figure~\ref{f:recordSahm}). The reason is that in 1934, the unemployment indicator peaked at 4pp but there was no recession. That peak is higher than many later recessionary peaks, and because of it, there are no rules based on the unemployment indicator that can have both no false positives (which would require the threshold to be above 4pp) and no false negatives (which would require the threshold to be below 1.5pp, the peak reached in the 1990 recession). 

Our recession rule, on the other hand, continues to work before World War 2, so it displays a perfect track record over 1929--2022. Using a threshold of 0.3pp to detect the starts of recessions, the minimum indicator perfectly identifies the 15 recessions of the April 1929--December 2022 period, without producing any false positive (figure~\ref{f:record1}). Over the entire April 1929--December 2022 period, on average, the minimum indicator detects recession starts with a delay of 1.5 months compared to the official recession starts determined by the \citet{NBER23}.

\section{Relation to other recession indicators}

Of course, there already exists numerous algorithms to identify the turning points of business cycles \citep{BB71,HP02,HP06,SW10a,SW14}, including in real time \citep{CP08,H11b,KLL23,L24}. These algorithms often use many different data series. For example, \citet{SW14} estimate US turning points based on a dataset of 270 monthly time series. 

However, \citet{CGL20a} finds that among all data available, labor-market data are the most reliable to date business cycles because they are less noisy and have fewer false positives. Another advantage of labor-market data is that they are less sensitive to revisions than GDP, so their real-time performance is almost as good as their final performance. 

\citet{CGL20b} adds that the unemployment rate, combined with a threshold rule, has a great record of identifying the beginning of recessions in the postwar US economy. This explains the long history, current popularity, and overall good performance of rules of that sort, such as the rule proposed by \citet[pp. 77--79]{S19}. Another rule of that sort is proposed by \citet[p. 159]{S08}: it compares the unemployment rate to its cyclical low (determined by hand) and uses a threshold of 0.4pp. Another similar rule was developed by \citet[pp. 118--120]{HS12a}: that rule compares the unemployment rate to its cyclical low (determined by hand) and uses a threshold of 0.35pp. Policymakers have also been using such rules. For instance, \citet[appendix 2]{B06a} compares the unemployment rate to its value 4 quarters earlier and uses a threshold of 0.3pp. Similar rules have been used in the private sector too. Goldman Sachs compares the unemployment rate to its cyclical low (determined by hand) and uses a threshold of 0.33pp \citep{S24}. A related private-sector rule is the Joshi rule developed by BCA Research \citep[p. 6]{P24}. The Joshi rule does not look at all unemployed workers but instead focuses on job losers not on temporary layoff. It also reduces the recession threshold to 0.2pp.

\citet{P24} recently confirmed \citeauthor{CGL20b}'s insight. \citeauthor{P24} attempts to improve the performance of the Sahm rule by using the unemployment rate jointly with the slope of the yield curve---a popular recession indicator developed by \citet{H88a,H89a}. However, using the yield curve does not much at all: \citet[p. 1]{P24} reports that ``for reasons I do not understand, it appears that the almost all available information about the state of the economy is encoded in the unemployment rate.''

Given the good performance of recession indicators based on the unemployment rate, it is unsurprising that our indicator does well. We are able to improve upon unemployment-only indicators by leveraging two insights from the macroeconomics of slack. First, business cycles are mostly driven by shocks to aggregate demand, which trigger shocks to labor demand \citep{MS15}. Therefore, recessions are mostly caused by drops in aggregate demand. Second, such shocks produce negative comovements between unemployment rate and vacancy rate as the economy moves along the Beveridge curve \citep{MS15,MS22,MS24b}. Therefore, a recession generally features both a drop in vacancy rate and a rise in unemployment rate. 

By looking at both unemployment and vacancy data, our indicator therefore provides a less noisy and more reliable signal of recessions than unemployment-only indicators. Our indicator takes more time to rise at the onset of recessions because it requires both unemployment to rise and vacancies to fall. Yet, maybe surprisingly, our recession rule provides a more rapid recession signal because the indicator on which it is based is less noisy.

In particular, our recession rule will not be triggered by small or large shifts in the Beveridge curve, which occur every so often \citep[figure 5]{MS21b}. It will not be triggered by an outward shift in the Beveridge curve because such shift produces a joint increase in unemployment and vacancy rate---so the vacancy rate does not fall. It will not be triggered either by an inward shift in the Beveridge curve because such shift produces a joint decrease in unemployment and vacancy rates---so the unemployment rate does not rise. Only a diminution of economic activity pushing the economy down the Beveridge curve will trigger the indicator.

\section{Conclusion}

This note constructs a new recession rule for the US economy by combining data on job vacancies and unemployment. The new rule is triggered earlier than the Sahm rule: on average it detects recessions 0.8 month after they have started, while the Sahm rule detects them 2.1 months after their start. The new rule also has a better historical track record: it perfectly identifies all recessions since 1929, while the Sahm rule breaks down before 1960. The rule indicates that the US economy may have entered a recession as early as March 2024. In August 2024, the probability that the US economy is in recession is 48\%.

Of course, it is useful to know in real time that the economy might has entered a recession. A recession indicates that economic conditions will rapidly deteriorate and calls for policy actions. In that way, our indicator conveys useful information for policymakers. But there is no direct link between the value of the indicator and good monetary and fiscal policy.

However, the combination of vacancy and unemployment data that we use to construct our recession rule can also be used to design good stabilization policies. This paper provides an example of the predictive power of the vacancy-unemployment combination. But the vacancy-unemployment combination has normative power too. 

Using the same data as in this paper, \citet{MS24} argue that the full-employment rate of unemployment (FERU) in the United States is given by $u^* = \sqrt{uv}$, where $u$ is the unemployment rate and $v$ the vacancy rate. The FERU is a central target for the federal government and Federal Reserve, because both are legally mandated to maintain the economy at full employment \citep{EA,FRRA,FEBGA}. Because the FERU is the efficient unemployment rate, it is a key input into the design of optimal monetary policy \citep{MS22,MS24b} and optimal fiscal policy \citep{MS19}.

Between 1930 and 2024, the FERU is stable, hovering around 4\% \citet[figure 12A]{MS24}. The FERU has generally been below the unemployment rate, so the US economy has generally fallen short of full employment \citet[figure 12B]{MS24}. As of August 2024, the unemployment rate is 4.2\%, the vacancy rate is 4.6\%, so the FERU is $u^* = \sqrt{0.042 \times 0.046} = 4.4\%$. Since the unemployment rate is below the FERU, the US labor market is still inefficiently tight. However, the unemployment gap is almost back to zero, at $u - u^* = 4.2\% - 4.4\% = -0.2$pp. Thus, the US economy is almost back at full employment after having been overheated for 3 years, since the middle of 2021. 

If the economy keeps cooling as it would in a recession, the labor market will rapidly cool past full employment and become inefficiently slack. In such situations, the Fed should cut rates to stimulate aggregate demand and labor demand. Rate cuts take some time to become fully effective \citep{C12}. Yet, they are the most natural way to keep the economy as close as possible to full employment.

\bibliography{\bib}

\end{document}